\documentclass[aps,amsfonts,amsmath,prd,preprint,nofootinbib]{revtex4}
\pdfoutput=1
\newcommand{\beq}{\begin{equation}}
\newcommand{\eeq}{\end{equation}}
\usepackage{graphicx}
\usepackage{natbib}

\usepackage{xcolor}

\usepackage{amsmath}

\newcommand{\be}{\begin{equation}}
\newcommand{\ee}{\end{equation}}
\newcommand{\bea}{\begin{eqnarray}}
\newcommand{\eea}{\end{eqnarray}}
\newcommand{\bi}{\begin{itemize}}
\newcommand{\ei}{\end{itemize}}

\begin{document}

\title{Parametric Resonances in Axionic Cosmic Strings}

\author{Jose J. Blanco-Pillado{$^{1,2,3}$}\footnote{josejuan.blanco@ehu.eus}, Daniel Jim\'enez-Aguilar{$^{1,2}$}\footnote{daniel.jimenez@ehu.eus}, Jose M. Queiruga{$^{4,5}$}\footnote{xose.queiruga@usal.es} and Jon Urrestilla{$^{1,2}$} \footnote{jon.urrestilla@ehu.eus}}

\affiliation{$^1$ Department of Physics, University of Basque Country, UPV/EHU, 48080, Bilbao, Spain \\
$^2$ EHU Quantum Center, University of Basque Country, UPV/EHU\\
$^3$ IKERBASQUE, Basque Foundation for Science, 48011, Bilbao, Spain \\
$^4$ Department of Applied Mathematics,
University of Salamanca, 37008, Salamanca, Spain \\
$^5$
Institute of Fundamental Physics and Mathematics,
University of Salamanca, 37008 Salamanca, Spain.
}

\begin{abstract}
In this letter we uncover a new parametric resonance of axionic cosmic strings.
This process is triggered by the presence on the string of internal mode excitations
that resonantly amplify the amplitude of transverse displacements of the string. 
We study this process by running numerical simulations that demonstrate
the existence of this phenomenon in a $(3+1)$ dimensional lattice field theory and
compare the results with the analytic expectations for the effective Lagrangian
of the amplitude of these modes and their interactions. Finally, we also analyze the
massless and massive radiation produced by these excited strings and
comment on its relevance for the interpretation of the results of current numerical 
simulations of axionic cosmic string networks.

\end{abstract}

\maketitle

\section{Introduction}

The nature of dark matter remains as one of the biggest puzzles in cosmology.
Among the multiple hypothetical particles that have been proposed as dark matter candidates, the
axion is perhaps one of the most promising ones. Particle physics models
with axions appear in many well motivated extensions of the Standard Model where an
extra $U(1)$ symmetry (the so-called Peccei-Quinn symmetry \cite{PQ}) is added. This symmetry is broken at high energies leading to the 
appearance of a Goldstone mode. Eventually, this mode becomes massive due to small instanton contributions
at low energies. It is therefore clear that these models lead to the prediction of a new particle with very weak 
interactions, in other words, a perfect candidate for dark matter \cite{Weinberg78,Wilczek78,Preskill1983,Abbot1983,Dine1983}.

In a cosmological context, one can assume that the Peccei-Quinn symmetry is spontaneously
broken before inflation. This would lead to an axionic abundance today that depends 
on the value of the axion field in our local patch of the universe. We will not consider these
models in this paper any further. On the other hand,
if this symmetry is broken after inflation, the phase transition would lead to the formation
of axionic cosmic strings \cite{Kibble:1976sj,Vilenkin:1982ks,Vilenkin:1986ku}. This cosmological network of strings will evolve in an expanding
universe producing a spectrum of Goldstone modes. This process will continue until the 
axion acquires its mass, leading to the formation of domain walls attached to the strings.
These walls will typically trigger the annihilation of the string network, a process that will also
contribute to the axionic abundance \footnote{The cosmological scenario after
the walls are formed strongly depends on the details of the axion potential \cite{Kawasaki:2014sqa}.}. 

It is clear from this description that in order to estimate the relic abundance
of axions in these models one needs to have a good understanding of the evolution of this
network of axionic strings. This task has been tackled by several groups over the years 
\cite{Sikivie:1982qv,Hagmann:1990mj,Yamaguchi:1998gx,Hagmann:2000ja,Yamaguchi:2002sh, Fleury:2015aca,Fleury:2016xrz,Klaer:2017qhr,Kawasaki:2018bzv,Gorghetto:2018myk,Vaquero:2018tib,Martins:2018dqg,Hindmarsh:2019csc,Gorghetto:2020qws,Hindmarsh:2021vih,Hindmarsh:2021zkt,Buschmann:2021sdq}.
However, the results of the different groups are not all consistent
with one another. One of the issues that is currently under scrutiny
is the large scale properties of the network, in particular the density of 
strings in the so-called {\it scaling regime}. This is, of course, a key element of the network that directly
affects the estimate of the axion abundance in the model.

Another important ingredient needed to compute the cosmological
density of axions today is the spectrum of Goldstone mode radiation produced
by strings. This spectrum has also been estimated by several groups using 
different techniques, although a quantitative agreement between different
groups has not been reached yet. In this paper, we would like to describe a phenomenon that may have
some relevance in the dynamics of axionic cosmic strings and in turn in the
spectrum of the radiation they produce.

As we mentioned above, these axionic cosmic strings appear as solitonic objects
in a field theory model with a $U(1)$ global symmetry. The study of small 
perturbations around the simplest straight string configuration demonstrates
the existence of two different types of possible excitations of these objects.
The lowest energy excitations correspond to the transverse displacements
of the strings. These perturbations can be viewed as massless modes on the 
$1+1$ dimensional worldsheet of the string and describe wiggles 
that propagate at the speed of light. Apart from these perturbations,
there is another type of excitations that have to do with the internal dynamics of the
soliton. These modes appear generically in many field theory models with 
solitons \cite{Arodz:1991ws, Goodband:1995rt,Kojo:2007bk,Alonso-Izquierdo:2015tta, Alonso-Izquierdo:2016pcq,Blanco-Pillado:2020smt, Blanco-Pillado:2021jad}. 
Physically, they represent the possible deformations of the shape of the relaxed soliton, which
means that their masses are typically similar to the energy scale associated to the size of the
soliton. However, an interesting aspect of these states is their long lifetime \cite{Manton:1996ex,Blanco-Pillado:2020smt,Blanco-Pillado:2021jad},
which could make them relevant for the long term dynamics of cosmological
defects.

The dynamics of global strings has been extensively studied in the literature 
within the thin wall approximation \cite{Vilenkin:1986ku,Garfinkle:1988yi,Davis:1988rw,Dabholkar:1989ju,Battye:1993jv,Battye:1995hw,Drew:2019mzc}. 
However, this effective action does not take into
account the presence of these massive internal modes of the string. Here we would
like to consider the effect that the presence of both types of excitations can 
have on strings. In particular, we will show that an internal mode
excitation could trigger the resonant amplification of the transverse motion of
the string. This, in turn, could also accelerate the decay process for the extra
energy stored in the string. This is an analogous mechanism to the one
recently discovered in domain wall strings in \cite{Blanco-Pillado:2022rad}. 

The organization of the rest of the paper is the following. In section II we
discuss the field theory model that we will consider as well as the
properties of the linear perturbations around the relaxed axionic string.
In section III we study an effective Lagrangian that describes the
coupling between these modes and investigate the possible appearance
of parametric resonances. In section IV we show how these resonances
are easily activated in lattice field theory simulations and compare the
results with the analytic predictions obtained in the previous section.
Finally, we end with some conclusions about the possible relevance
of these findings to the results obtained in field theory simulations of
cosmic string networks.

\section{Axionic Strings and their Excitations}

We will consider the following model for a complex scalar field with 
a $U(1)_{PQ}$ invariant action of the form
\beq
\label{action}
S = \int{ d^4x \left[\partial_{\mu} \phi^{*}   \partial^{\mu} \phi - \frac{\lambda}{4} \left(\phi^* \phi - \eta^2\right)^2\right]}\,,
\eeq
where $\lambda$ describes the quartic coupling of the field and $\eta$ denotes
the energy scale of the theory. The form of the potential leads to a symmetry breaking where 
the vacuum manifold is parametrized by the configurations with $|\phi(x)| = \eta$. The spectrum of fluctuations around these vacua is composed
of a massive (radial) field whose mass is $m_r = \sqrt{\lambda} \eta$ and a 
massless Goldstone mode that represents the perturbations in the field angular
phase. The equations of motion obtained from the Lagrangian are
\beq
\partial_{\mu} \partial^{\mu} \phi + \frac{\lambda}{2} \left(\phi^* \phi - \eta^2\right) \phi = 0~.
\label{eq:eoms}
\eeq

Since the first homotopy group of the vacuum manifold is non-trivial, we will look for
static string-like solitonic solutions of the
form
\beq
\phi_v(x)= \eta f(\rho) e^{i \theta}\,,
\eeq
where we have used the cylindrical coordinates $(\rho, \theta)$ to parametrize the plane
transverse to the string. With this ansatz, one can find the equation of motion for
the profile function $f(r)$, namely
\beq\label{global-field-eq}
\frac{d^2 f}{dr^2}+\frac{1}{r}\frac{df}{dr}-\frac{1}{r^2} f-\frac{1}{2}\left(f^2-1\right)f=0~,
\eeq
where we have introduced the dimensionless radial
coordinate $r = m_r \rho$. The solution of this equation describes a solitonic object
with an inner region with $f(r) \approx c~ r + ...$ where the field is close to the top of the
potential; and an asymptotic region where the field approaches the vacuum as $f(r) \approx 1 - (1/r^2) + {\cal O}(1/r^4)$ (see 
\cite{Blanco-Pillado:2021jad} for more details on this solution).

It is important to realize that this solution has a transverse energy density that falls
slowly with the distance from the core. This leads to a logarithmic divergence for
the energy per unit length of the string. This is a manifestation of another important
point. These strings are charged with respect to the Goldstone mode of the theory,
and this divergence is just an indication of this fact. Incorporating this coupling to the
effective theory of the string means that one should not only take into account the
Nambu-Goto action, but also consider the contribution of the Kalb-Rammond term \cite{Kalb:1974yc}.

This divergence does not present a difficulty cosmologically since its contribution is
always cut-off by the presence of another string at some distance $R$ from the original
string. This leads to an energy per unit length of the form
\beq
\mu (R)\approx \mu_{\text{core}} + 2 \pi \eta^2 \log \left(\frac{R}{\delta}\right)~,
\eeq
where $\delta$ describes a measure of the size of the inner core of the solution.
In a lattice field theory simulation for a string network, the distance between the
strings is obviously much smaller than in a realistic cosmological setting. This
could have an important effect on the dynamics of these strings,
leading, for example, to transient effects in the simulations.

\subsection{String excitations}

Let us now consider the possible perturbations around the static
straight axionic string presented earlier. In order to study these
fluctuations, we take the ansatz
\beq
\phi({\bf x},t) = \phi_v({\bf x}) + \delta \phi({\bf x}) \cos(w t)~.
\eeq
We can now decompose the perturbations in waves
along the longitudinal direction of the string, the $z$
direction, of the form
\beq
\delta \phi({\bf x}) = A ~\delta \phi (r, \theta) \cos (w_z  z)~.
\eeq

This decomposition leads to a linear system of equations that
formally looks like a Schr\"odinger equation for the function  $\delta \phi (r, \theta)$ 
in a radial potential whose eigenvalue is given by the combination $w^2 - w_z^2$.
This eigenvalue problem is identical to the linear equations for
perturbations of the $2d$ vortex, which has been recently studied
in detail in \cite{Blanco-Pillado:2021jad}. We can use the results
obtained in that paper to identify the lowest energy states of these
fluctuations in the $3+1$ string configuration.

We first note that there are zero mode solutions of these equations
with $w^2 - w_z^2=0$. These modes correspond to travelling
waves moving at the speed of light along the longitudinal 
direction of the string. Physically, they represent the Goldstone
modes describing the fact that the static string breaks the translational
invariance of the theory. When promoted to $3+1$ dimensions, these
wiggles can be viewed as zero mass excitations on the string worldsheet.

These zero modes are easy to compute since they are just given by
the derivative of the static solution along a direction perpendicular to the
string. In particular,
we have
\beq
\eta^{0}_x (r, \theta) \equiv \partial_x \phi_v ({\bf x})~~~~~~~~~~\text{and}~~~~~~~~~~~~\eta^{0}_y(r, \theta) \equiv \partial_y \phi_v ({\bf x})\,.
\eeq

Note that this zero mode is not normalizable. However, this is just
another manifestation of the slow decline of the energy density
of the global strings. One can, in fact, find an exact solution of the
non-linear equations of motion that represents an arbitrary transverse wiggle
moving at the speed of light \cite{Vachaspati:1990sk}. These zero modes
are therefore the linear approximation of those exact solutions.

\begin{figure}[h!]
\includegraphics[width=12cm]{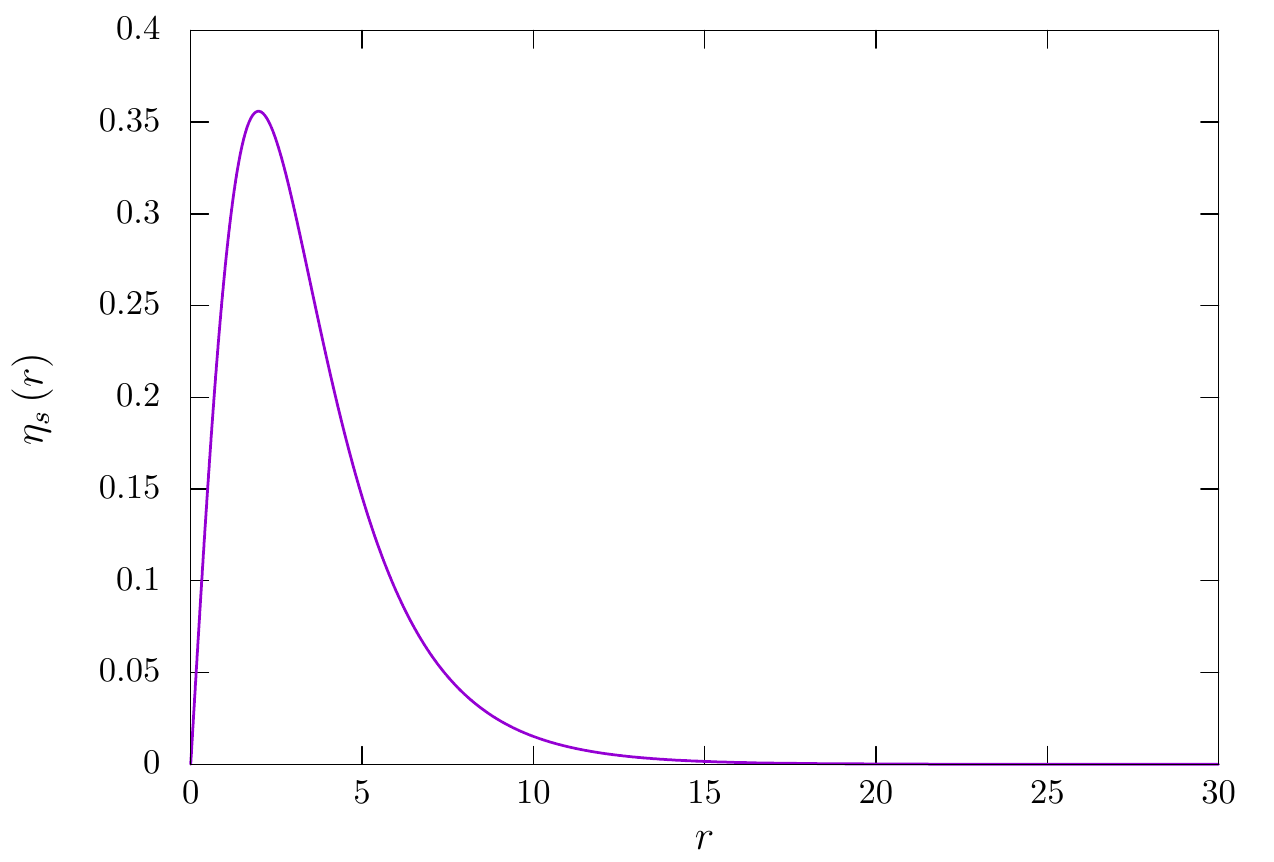}
\caption{Bound state function for the first massive mode.} 
\label{fig:perturbations-modes}
\end{figure}

Furthermore, it was shown in \cite{Blanco-Pillado:2021jad} that this
eigenvalue problem possesses an infinite number of bound states in the radial
part of the scalar field. These bound states can be found numerically
and they represent certain deformations of the internal shape of the
vortex. In particular, the first two such modes were obtained numerically
in  \cite{Blanco-Pillado:2021jad}. Here we just reproduce in Fig. \ref{fig:perturbations-modes}
the first mode $\delta \phi (r, \theta) = \eta_s(r) e^{i \theta}$,
which will be the focus of our study. The eigenvalue in this case
was found to be $w_{s}^{2}\equiv w^2 - w_z^2=0.8133$ in units of the square of the mass $m_{r}$.

Finally, apart from these bound states we have the continuous
scattering states that correspond to the perturbative fluctuations
that propagate in the vacuum. These are the massless Goldstone modes
that represent fluctuations on the phase of the scalar field and
the massive radial excitations whose mass starts at $m=1$ in these
units.

As it was shown in  \cite{Blanco-Pillado:2021jad}, an excited vortex
state given by a small amplitude of the {\it shape mode}, $\eta_s(r)$, 
could take a long time to decay to its ground state. The reason
for this is that its frequency is below the mass threshold for
radial modes to propagate in the vacuum. This does not mean,
however, that this state is stable since non-linear interactions
allow this energy to slowly leak to infinity. This phenomenon
can be directly translated to our $(3+1)$ configuration by
considering a homogenous excitation of the shape mode
along the string. The invariance along the $z$ direction
of this configuration allows us to use the results from the
$2+1$ problem. 

The situation becomes more interesting in $(3+1)$
dimensions due to the non-linear interaction between the
different excitations on the string. As we will see in the next
sections, we will encounter several types of resonances
in the system of interacting modes.

\section{Parametric Resonances}

Let us now consider the situation of a straight axionic string
uniformly excited with the presence of a shape mode. As 
we described earlier, this mode will couple non-linearly to
radiation that slowly leaks this extra energy to infinity in the 
direction perpendicular to the string. This was already studied 
in  \cite{Blanco-Pillado:2021jad}. Here we will be interested
in studying the coupling between the shape mode and
the transverse wiggle excitations of the string. With this idea in mind,
we consider a possible excitation of the string of the form
\beq
\phi(t,{\bf x}) = f(r) e^{i \theta} + A(t) \eta_s(r)e^{i \theta} + D(t) \eta^{0}_x(r,\theta) \cos(w_z z)\,,
\label{eq:ansatz resonance}
\eeq
where $A(t)$ denotes the amplitude of the shape
mode and we take $D(t)$ to represent the amplitude
of a standing wave of the displacement of the string
along the $x$ direction\footnote{Note that this ansatz disregards completely the
presence of any radiation field. This is of course an approximation. Taking this into consideration
would become important when we compare the results of our numerical simulations with the
predictions of this ansatz.}$^{,}$\footnote{ Note that this effective model can be easily extended to describe displacements in the $xy$ plane taking into account that any zero mode perturbation of the string can be expressed as a linear combination of the zero modes in the $x,y$ directions.} . We can now insert this ansatz 
in our action in Eq. (\ref{action}) to obtain, after integrating
along the spatial directions, an effective Lagrangian
of the form

\bea\label{effec_Lag}
L_{\text{effec}} = - \mu(R) L_z  + 2 \pi \eta^2 L_z &\Big[&{\dot A}^2 - w_s^2  A^2 - I_{A,3} A^3 -  I_{A,4}  A^4\nonumber\\
&+& I_{D} (R) {\dot D}^2 - (I_{D,2} + I_{D}(R) w_z^2 ) D^2 -  I_{D,4}  D^4\nonumber\\
&-& I_{AD,3} ~A D^2 - I_{AD,4} ~A^2 D^2 \Big]\,,
\eea
where the coefficients $I_{A,3}, I_{A,4}, I_{D,2}$ and  $I_{D,4}$
are computed by performing the integrals specified in the Appendix A and
represent the finite higher order couplings between these amplitudes. On the other hand, 
the coefficient $I_{D}(R)$ can be shown to be logarithmically divergent with the cutoff that
we implemented at large distances, namely the distance $R$.

The equations of motion we obtain from this effective Lagrangian are given by
\bea
\label{full-analytic-eqs}
&&\ddot A(t) + \left( w_s^2 +  I_{AD,4} ~ D(t)^2 \right) A(t) + \frac{3}{2} I_{A,3}   A(t)^2 +   2 I_{A,4} A(t)^3 +\frac{1}{2}   I_{AD,3}  D(t)^2 =0~,\nonumber\\ \nonumber\\
&&\ddot D(t) + \left( w_z^2 +   \frac{I_{D,2}}{I_D(R)} +   \frac{I_{AD,3}}{I_D(R)} A(t) +  \frac{I_{AD,4}}{I_D(R)} A(t)^2  \right) D(t) + 2 \frac{I_{D,4}}{I_D(R)} D(t)^3  =0 ~.
\eea

At the lowest order, these equations recover the information about
the shape and zero modes that we already discussed in the study of the
linear regime\footnote{Note, however, that the zero mode acquires a small
mass term due to a finite cutoff scale $R$. Taking the limit of $R\rightarrow \infty$,
this term will vanish.}. Going beyond the linear order, we identify the presence
of resonance effects between these two modes. Let us start by looking at
the zero mode equation. Disregarding the higher order coupling, we get 
an equation of the form
\beq
\label{Mathieu-for-D}
\ddot D(t) + \left( w_z^2 +   \frac{I_{D,2}}{I_D(R)} +   \frac{I_{AD,3}}{I_D(R)} A(t)  \right) D(t)   =0 ~.
\eeq
Assuming the unperturbed shape mode time dependence, namely $A(t) \propto \cos(w_s t)$,
we immediately notice that the form of the equation for the displacement of the
string is of the Mathieu type \cite{MathieuEq}. This means that we should expect a parametric
resonance behaviour of the string zero mode in the presence of a shape mode
excitation. Furthermore, using the information about the different bands of instability
of the generic Mathieu equation, we can compute the condition for the zero mode 
wavelength where we should expect the resonant amplification. This condition
can be shown to be
\beq
\sqrt{w_z^2 +   \frac{I_{D,2}}{I_D(R)}} = \frac{w_s}{2}~.
\eeq

On the other hand, in situations where the amplitude of the
shape mode is small, its equation of motion becomes
\beq
\ddot A(t) +  w_s^2  A(t)  +\frac{1}{2}   I_{AD,3}  D(t)^2 =0\,,
\eeq
so we also find a possible source of resonance in case of an oscillating zero mode in $D(t)$.
This means that one can also transfer energy resonantly from the zero
mode to the internal shape mode, but only when the string oscillates at a
frequency comparable to $w_s$.

This analysis indicates that there are resonant effects in the
interaction between these modes. However, we do not expect any
runaway process since higher order terms will tame any dramatic
behaviour. In reality, we expect to have a transfer of energy between
these modes back and forth.\footnote{This was indeed already observed in a similar
situation in another field theory model of solitons in $2+1$ dimensions in \cite{Blanco-Pillado:2022rad}.}
This, however, will have an effect
on the amount of radiation emitted from the string which is not
taken into account in this effective Lagrangian. We will comment
more on this in the next section.

\section{Numerical Simulations}

The analysis presented in the previous section suggests that a
uniform excitation of the shape mode on a string will induce a
parametric excitation of a zero mode of a particular wavelength.
In this section we show this is indeed the case by running a
lattice field simulation of our model.\footnote{We move the details
of the implementation of this numerical simulation to  Appendix B.}

 \begin{figure}[htb]

 \includegraphics[width=7.0cm, height=0.8cm]{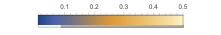}

\includegraphics[width=4.2cm, height=6.3cm]{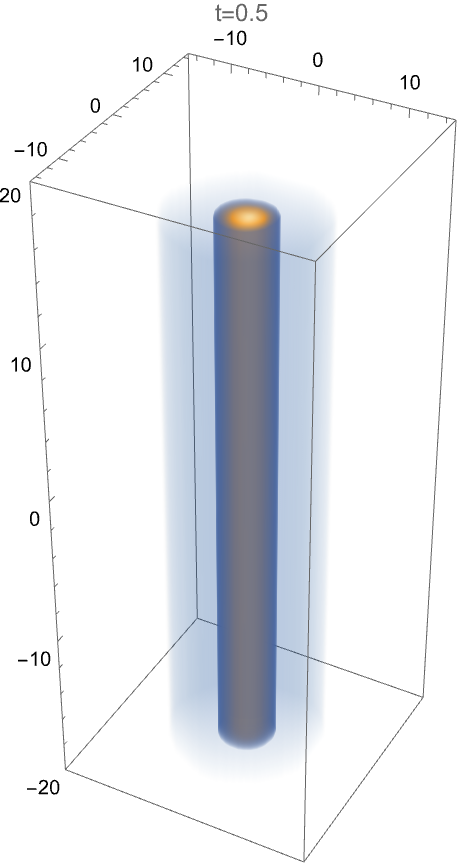}
\includegraphics[width=4.2cm, height=6.3cm]{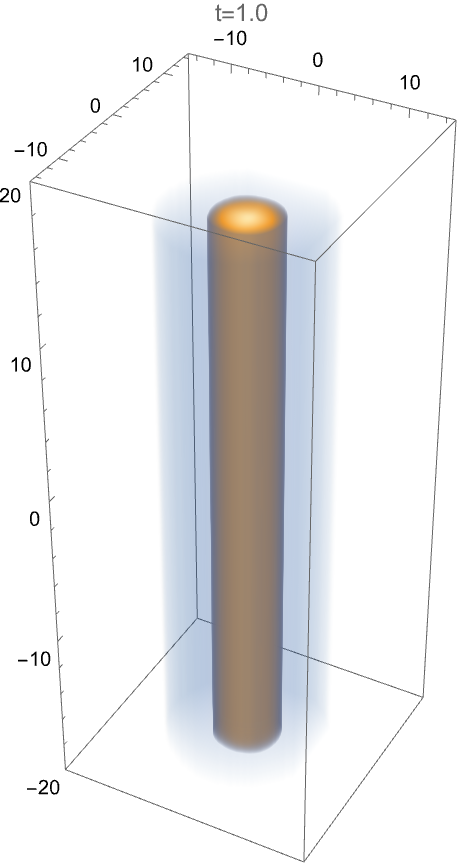}
\includegraphics[width=4.2cm, height=6.3cm]{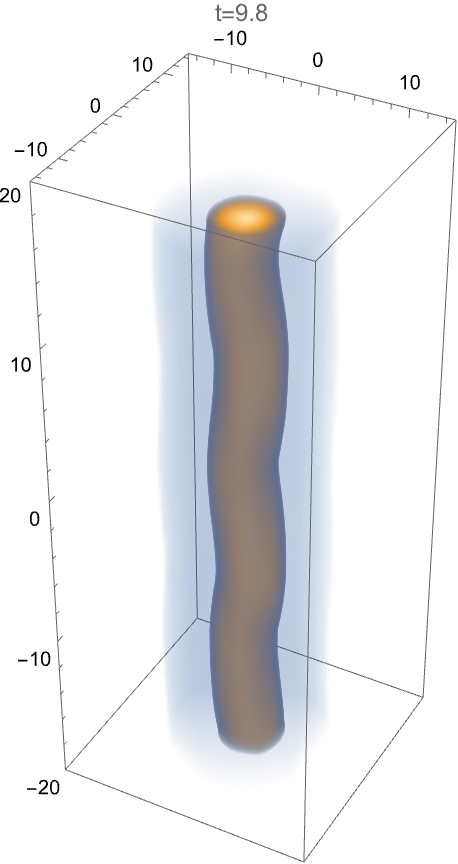}
\includegraphics[width=4.2cm, height=6.3cm]{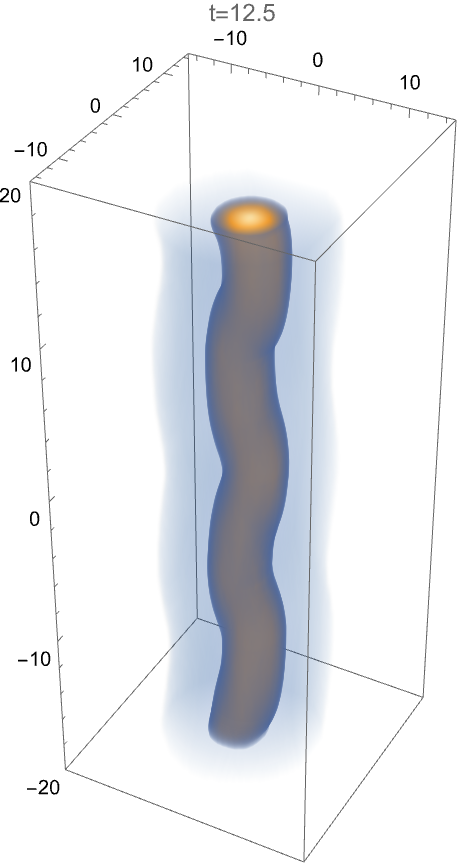}
\includegraphics[width=4.2cm, height=6.3cm]{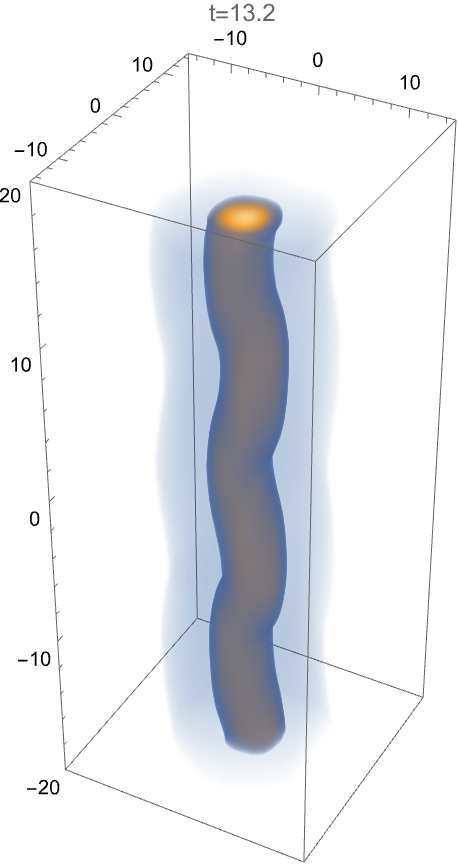}
\includegraphics[width=4.2cm, height=6.3cm]{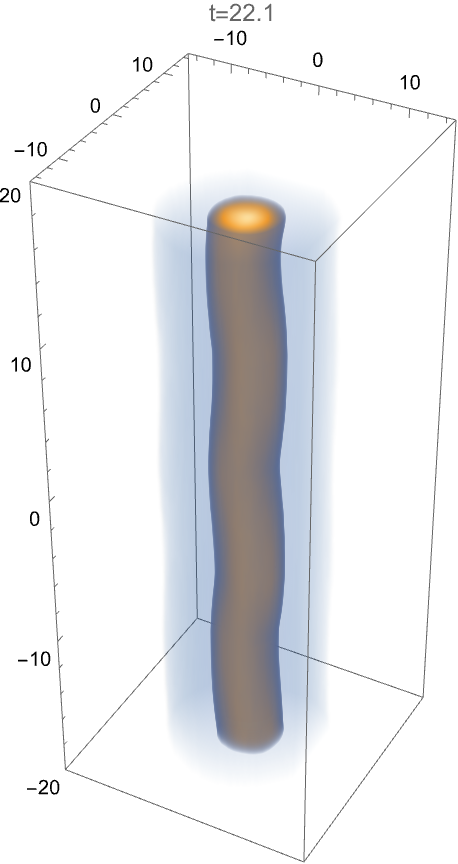}
\caption{Evolution of the string energy density in field theory. The initial state corresponds to an
excited state with $A(t=0)=0.5$ and $D(t=0)=0.01$. Top panel, left-to-right, $t=0.5,\, 1.0$ and $9.8$. Bottom panel, left-to-right $t=12.5, \, 13.2 $ and $22.1$. The color palette indicates energy density in units of $\lambda\eta^4$.}
\label{fig:parametric-resonance-from-field-theory}
\end{figure}

\begin{figure}[h!]
\includegraphics[width=8cm]{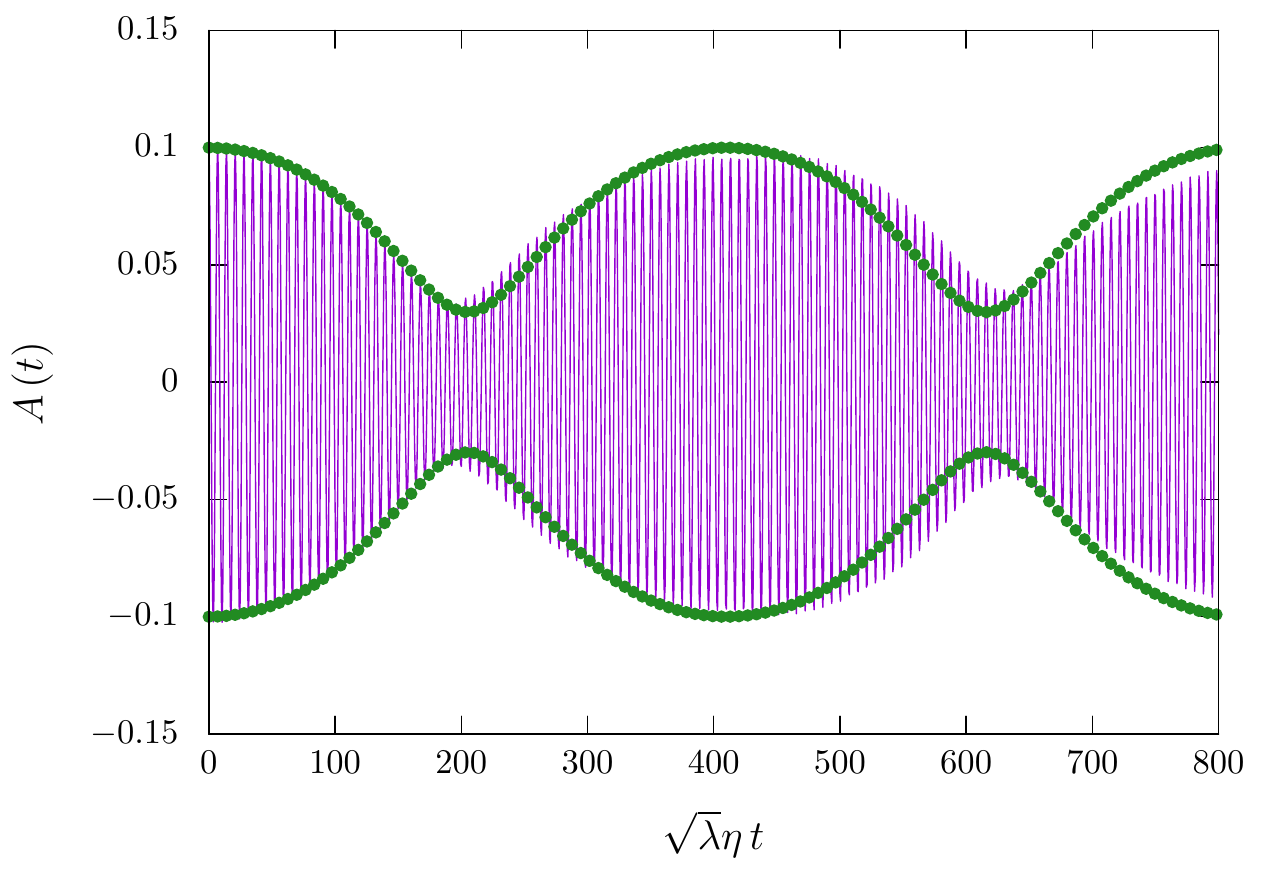}\includegraphics[width=8cm]{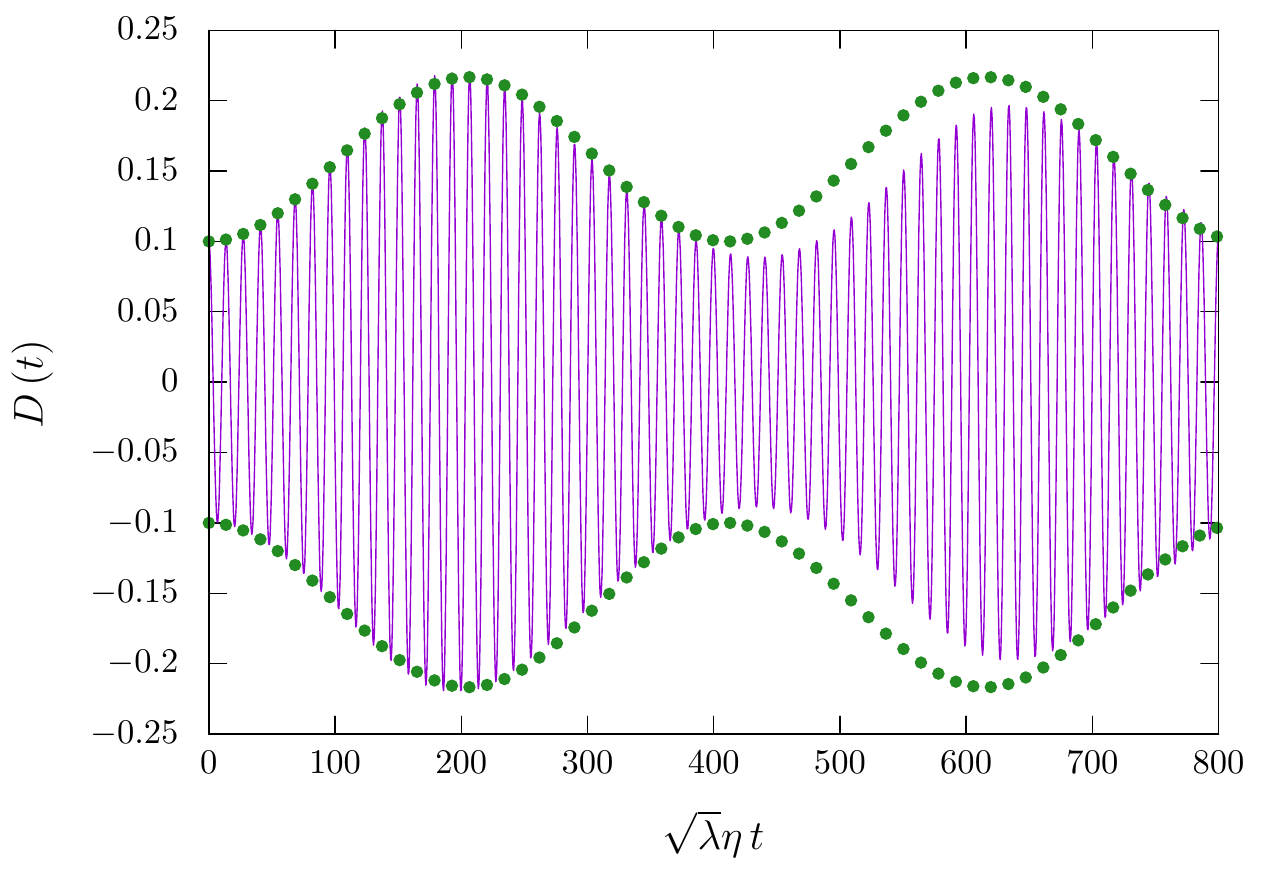}
\caption{We show the comparison of the measurements of the shape mode (left) and zero mode
(right) from the field theory simulation (in purple) with the results predicted using the equations
of motion of our effective Lagrangian (green).} 
\label{fig:comparison-theory-simulation}
\end{figure}

We show in Fig. \ref{fig:parametric-resonance-from-field-theory} a few snapshots of the position of the string
extracted from our field theory simulation starting from an excited 
state with a shape mode with amplitude $A(t=0)=0.5$. We notice how an
initially straight string develops a modulation whose amplitude
grows for a while before it starts decreasing again\footnote{In order to accelerate this
process in our simulations we give the string a tiny small initial modulation along the $x$ axis to break the
symmetry. We have also performed numerical experiments where we place the string in the presence of 
small random perturbations. This random initial state also triggers the onset of the instability although
it takes a longer time to appear.}. This is exactly what the analytic system of equations given in the previous section
predicts. The initial growth is representative of the parametric
resonant effect of the Mathieu type equation for the amplitude
of the zero mode in the presence of the internal excitation, namely
Eq. (\ref{Mathieu-for-D}). This behaviour is however cut-off by the presence of
non-linear interaction terms in the system of equations. In 
order to further test the analytic system, we extract the amplitude
of both the shape and zero mode directly from the simulation.
We can easily do that by projecting the solution onto those
modes at any moment in time.\footnote{We show how this is done  in Appendix B.} Following this procedure, we
can obtain Fig. \ref{fig:comparison-theory-simulation}, where we compare the predictions of the
system of equations given in Eqs. (\ref{full-analytic-eqs}) with the amplitudes of the different modes in
the field theory simulation. We note that the solution of the analytic equations is very sensitive to
the numerical values of the coefficients. We have found that in order to have a perfect
agreement with the results of our simulations we need to slightly modifiy the numerical
coefficients obtained for this setup. We believe this is due to the inaccuracies in the calculation
of the coefficients due to numerical error in the functions involved in these calculations (see Appendix A).

Using these coefficients we see a very good agreement with the theoretical expectation for the first oscillation. 
The subsequent slight deviation is also to be expected since the field theory simulation allows for the presence of radiation
from the string, something that is not included in our analytic treatment.

\subsection{Radiation from the excited string}

As we explained in the previous sections, an axionic string
is coupled to the massless Goldstone mode that propagates in the
vacuum. This can be easily seen if one excites a zero mode on the
string. The initial amplitude of this excitation will decrease as the
string oscillates mainly due to massless radiation. This has been
studied in the literature \cite{Vilenkin:1986ku,Davis:1988rw,Battye:1993jv,Battye:1995hw}, more notably recently with the
aid of adaptive mesh refinement techniques in \cite{Drew:2019mzc,Drew:2022iqz}. We have
also performed this type of simulations and obtained a similar
result. Most of the energy emitted from an initial zero mode 
excitation occurs in the form of massless modes assuming a
low amplitude regime. Increasing the amplitude of these
oscillations, one encounters a mix of massive and
massless radiation due to non-perturbative radiative 
processes similar to the ones observed in other
analogous situations in other soliton models (see for
example \cite{Olum:1999sg,Blanco-Pillado:2022rad}).

On the other hand, a string excited with a homogeneous 
shape mode perturbation will decay mostly in terms of massive
radiation. This was demonstrated in the context of $2+1$ 
dimensional vortices in \cite{Blanco-Pillado:2021jad}, where we were able to obtain
an analytic description of the slow decrease of the amplitude
of this mode as a function of time. We have corroborated that
this is the case in our current $(3+1)$d setup in cases where there is no
instability.

\begin{figure}[h!]
\includegraphics[width=11cm]{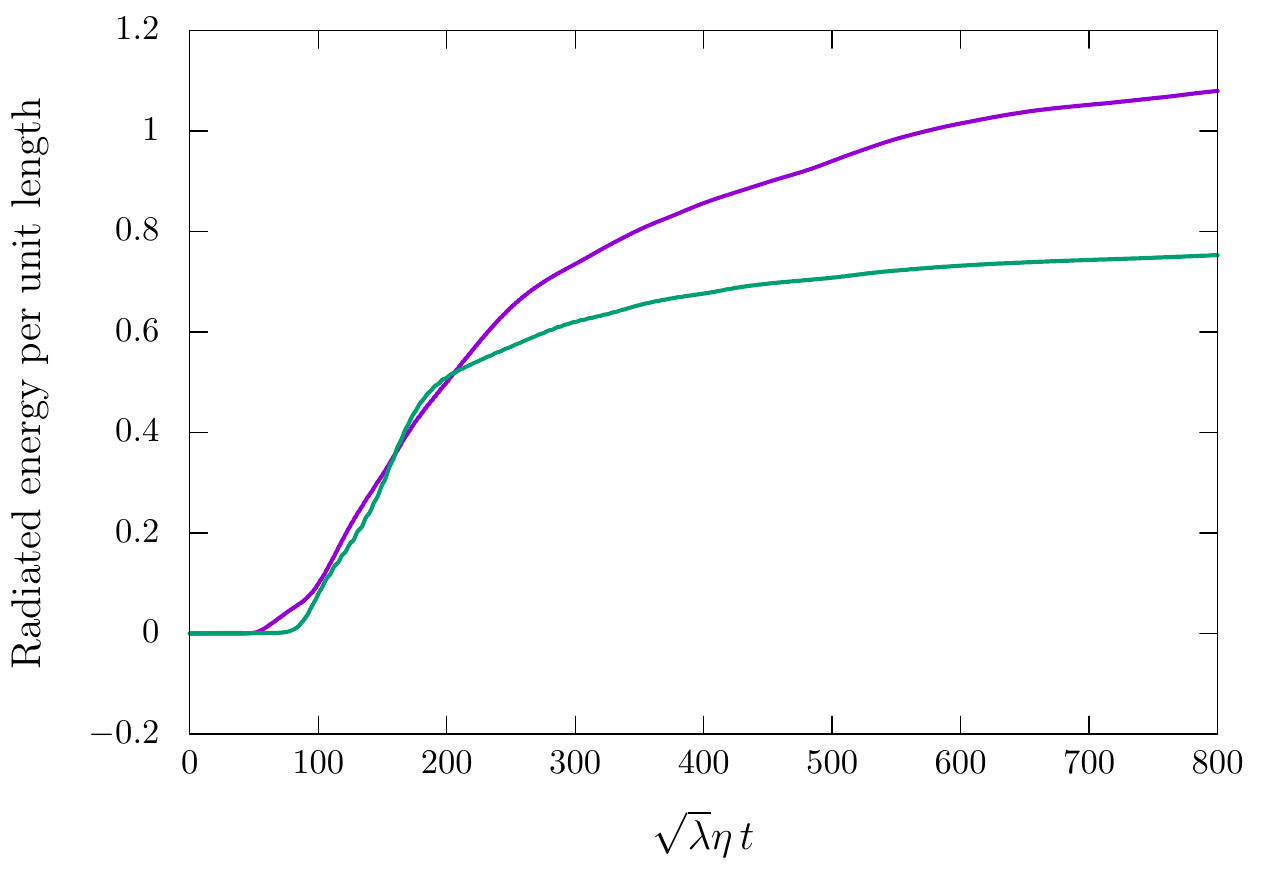}
\caption{Amount of energy radiated in the form of the massive mode (purple)
and the massless mode (green) as a function of time. The quantity shown in the $y$ axis is a dimensionless 
energy per unit length, namely, energy per unit length divided by $\eta^2$. In this simulation, the shape 
mode is initialized homogeneously with amplitude $A(t=0)=0.6$, which corresponds to an initial extra energy per unit length of $2.1$ in these units.}
\label{fig:power-emitted}
\end{figure}

The situation becomes more interesting in the presence
of the parametric resonance we discussed earlier. The initial
excitation of the shape mode leads to massive radiation due
to the coupling of this mode to the massive scattering states.
However, the subsequent amplification of the zero mode leads
to the appearance of a contribution of massless radiation. The non-linear
coupling of the shape and zero modes and their oscillating amplitudes
give rise to a complicated pattern of mixed radiation of both
modes, massless and massive radiation. We show in Fig. \ref{fig:power-emitted}
the different contributions of the total energy radiated by the string. We
compute this by integrating in time the power of each contribution  
going through a surface at a large distance from the string.\footnote{See the
detailed description of the expression of the power computed from
the simulation in Appendix B.}

\begin{figure}[h!]
\includegraphics[width=11cm]{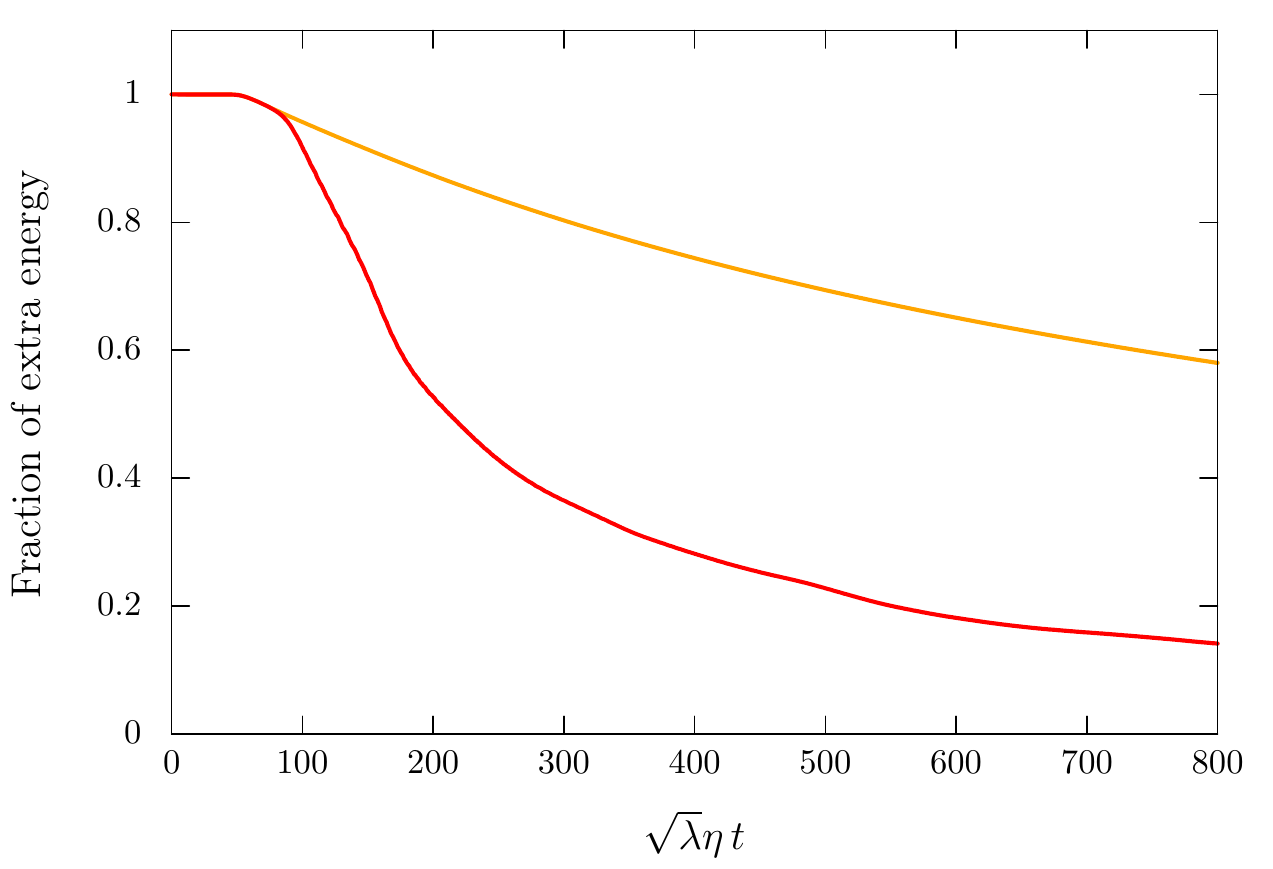}
\caption{Comparison of the extra energy for an excited string
with the same initial amplitude of the shape mode but different
size in the $z$ direction. In orange we show the fraction of this initial 
energy in the case of an excited string where the parametric resonance is prevented by the short length
in the $z$ direction. In red, we show the case where the resonance
is allowed and indeed it kicks in. Both these simulations use absorbing
boundary conditions for the ($x,y$) boundaries of the box.}
\label{fig:energy-decay}
\end{figure}

Finally, we notice that the total energy radiated from the string
is increased in the presence of a resonance. This means
that the energy stored in the initial internal excitation is radiated away 
faster than one would have previously estimated from the
$2+1$ calculation \cite{Blanco-Pillado:2021jad}. We show this 
effect in Fig. \ref{fig:energy-decay} by comparing the extra energy
in the simulation box for an excited string where the resonant
amplification is allowed versus the case where this possibility is
prohibited by a small $z$ direction. We see clearly that the 
presence of the zero mode excitation and the massless radiation
deplete the energy of the system faster. This is not surprising
since, as we mentioned, this excitation opens up a new channel
for decay by the coupling of the zero mode to the massless
Goldstone in the vacuum.

\section{Conclusions}

In this paper, we have shown that axionic strings can have parametric
resonant effects due to the interaction between massless
modes that parametrize the transverse motion of the string and
the massive shape modes that modify the internal structure of the soliton.
The presence of this instability leads to quantitative as well as qualitative differences
in the spectrum of radiation from axionic strings. This makes this study interesting
since it could have an effect in the estimate of axionic dark matter
abundance from cosmological global string networks.

In order for this effect to become important, the strings must be in an excited state
where part of the energy is stored in the shape mode. One of the instances where
this can happen is, of course, during the formation of the strings. This has been 
studied in a $(2+1)d$ context in \cite{Blanco-Pillado:2021jad}, where the results indicate that
the amount of extra energy is not very relevant. However,  it is 
important to remark that this conclusion may be different in a numerical simulation
depending on the way one produces the initial conditions. Some of the lattice
field theory simulations of these axionic strings use some period of relaxation
where these internal modes are presumably heavily suppressed. Others do 
not have this friction regime and could easily produce strings with an
initial extra energy in these modes. Using our results in this paper, one
can speculate that this would lead to a transient regime where massive
as well as massless radiation would be efficiently produced from the 
network. In particular, some simulations have found a significant amount
of massive radiation in the early stages of their evolution \cite{Gorghetto:2020qws}. It would be
interesting to understand whether this is due to the presence of these
internal modes on the simulated strings and whether the instability we
have discovered here is present.

Another way where this could be relevant is in the course of the network's
evolution whenever strings intercommute. In those moments, it is likely that
some of the energy could be absorbed by the string's internal modes. In fact,
this has been the claim put forward in \cite{Saurabh:2020pqe} based on the formation of loops
from the intersection of long strings. Interestingly, this paper also shows
that part of the spectrum of radiation from these loops is in the form 
of massive radiation, which would seem to be in agreement with our
observations. It would be interesting to analyse these simulations in detail
to identify whether or not there is any instability similar to the one 
presented in this paper.

The results we have obtained in this paper are quite similar
to the parametric resonance in domain wall strings found in \cite{Blanco-Pillado:2022rad}.
We argued there that this effect could be encapsulated by adding
to the effective action a term that represents the coupling of the
scalar field describing the internal mode and the Ricci scalar of the
string worldsheet. We conjecture that a similar term could be 
important for axionic strings and leave for future work to 
study the relevance of this term in an effective action for global strings 
which include a Nambu-Goto as well as a Kalb-Ramond terms.

Finally, it is clear that the mechanism that we described here would
also be present in other field theory models with solitons where there 
are both types of excitation modes living on the worldvolume of the defects. In particular, it would be
interesting to identify whether this can happen in the case of local 
strings or more generically in higher dimensional defects, like brane-like
objects. We leave this for future investigations.

\section{Acknowledgements}

We are grateful to Ken Olum, Tanmay Vachaspati, Alex Vilenkin and Giovanni Villadoro for 
many useful discussions and comments.  This work is supported in part by the PID2021-123703NB-C21 grant funded by MCIN/AEI/10.13039/501100011033/ and by ERDF; ``A way of making Europe''; by MCIN (project  PID2020-113406GB-I00), the Basque Government 
grant (IT-1628-22) and the Basque Foundation for Science (IKERBASQUE). The 
numerical work carried out in this paper has been possible thanks to the computing infrastructure 
of the ARINA cluster at the University of the Basque Country (UPV/EHU).

\appendix

\section{Effective Lagrangian integrals}

In this appendix we show explicitly the form of the space-dependent integrals in the effective Lagrangian (\ref{effec_Lag}). We impose the following normalization for the shape mode
\be
\int_0^\infty \,r\, dr \,\eta_s^{2}(r)=1.
\ee
The zero mode in the $x$-direction is decomposed as follows
\be
\eta_0^x(r)=\eta_1(r)+\eta_2 (r)e^{2i \theta},
\ee
where
\be
\eta_1(r)=\frac{1}{2}\left(f'(r)+\frac{f(r)}{r}\right),\quad \eta_2(r)=\frac{1}{2}\left(f'(r)-\frac{f(r)}{r}\right).
\ee
From the asymptotic form of $f(r)$, $f(r)=1-1/r^2+\mathcal{O}(1/r^4)$ it is easy to see that, for large $r$, the modulus of the zero mode grows as $1/r$, and as a consequence the zero mode is not normalizable. This is the origin of the logarithmic divergences in (\ref{effec_Lag}). The convergent integrals are given by the following expressions
\bea
I_{A,3}&=&\int_0^\infty  \,dr\, r \, f(r) \eta_s^3(r)\approx 0.197, \\
I_{A,4}&=&\frac{1}{4}\int_0^\infty  \,dr\, r \, \eta_s^4(r)\approx 0.0188, \\
I_{D,4}&=&\frac{3}{32}\int_0^\infty  \,dr\, r \, \left(\eta_1^4(r)+4\eta_1^2(r)\eta_2^2(r)+\eta_2^4(r)\right)\approx 7.36\times 10^{-4}, \\
I_{AD,3}&=&\int_0^\infty  \,dr\, r \,\eta_s(r)\left(\eta_1^2(r)+\eta_1(r)\eta_2(r)+\eta_2^2(r)\right)\approx 0.0904, \\
I_{AD,4}&=&\frac{1}{2}\int_0^\infty  \,dr\, r\, \eta^2_s(r)\left(\eta_1^2(r)+\eta_1(r)\eta_2(r)+\eta_2^2(r)\right)\approx 0.0185, \\
I_{D,2}&=&\int_0^\infty  \,dr\, r \, \left(\frac{1}{2} f^2(r)\left(\eta_1^2(r)+\eta_1(r)\eta_2(r)+\eta_2^2(r)\right)+\frac{1}{2} \left(\eta_1'^2(r)+\eta_2'^2(r)\right)\right.\\
&&\left. -\frac{1}{4} \left(\eta_1^2(r)+\eta_2^2(r)\right)+\frac{2}{r^2} \eta_2^2(r)\right)\approx -1.86\times 10^{-4}.\nonumber 
\eea
 
 The remaining, logarithmically divergent integral, has the form
 \be
 I_D(R)=\frac{1}{2}\int_0^R   \,dr\, r \, \left(\eta_1^2(r)+\eta_2^2(r)\right).
 \ee
For $R=40$, its numerical value is $I_{D}(40)\approx0.806$.\\\\ As we pointed out in the main text, complete agreement in Fig. \ref{fig:comparison-theory-simulation} is not obtained unless we slightly modify a few numbers involved in Eqs. (\ref{full-analytic-eqs}). In the example we have shown, the required changes are the following:
\begin{itemize}
\item $I_{AD,3}\,\,\rightarrow\,\,1.17\times I_{AD,3}$
\item $w_z^2+I_{D,2}/I_D(R) \approx 0.207 \,\,\rightarrow\,\,0.2077$
\end{itemize}
We attribute these imprecisions to the fact that all the functions involved in the integrals above are found numerically, and thus they come with some degree of uncertainty. The integral $I_{D,2}$ was found to be particularly sensitive to the number of points used for the integrand.

\section{Implementation of our numerical lattice field theory simulations}
In this appendix we provide the details of the lattice field theory simulations discussed in the main text. The simulations were run in a $3+1$ dimensional box with sides of lengths $L_{x}$, $L_{y}$ and $L_{z}$ where the string lied along the $z$ direction. The lattice spacing $\Delta x$ was chosen to be the same in the three directions, so the total number of lattice points was $L_{x}\times L_{y}\times L_{z}\,/\left(\Delta x\right)^{3}$. Finally, $\Delta t$ will denote the time step used in the simulations. \\\\
We solve the equations of motion employing the staggered leapfrog method and nearest neighbours for the discretization of the derivatives. Moreover, we implement Message Passing Interface (MPI) to handle the huge number of lattice points, which is typically of several million.

\subsection{Dimensionless variables}
In order to solve the equations of motion  (\ref{eq:eoms}) in a discrete lattice, we first define the following dimensionless variables:
\begin{equation}
\tilde{x}^{\mu}=m_{r} x^{\mu}, \,\,\,\,\, \tilde{L}_{x,y,z}=m_{r} L_{x,y,z}, \,\,\,\,\, \tilde{\phi}=\phi/\eta\,, 
\label{eq:dimensionless variables}
\end{equation}
where $m_{r}=\sqrt{\lambda}\eta$ is the mass of small fluctuations of the radial part of the field about the vacuum. With these redefinitions, the action reads
\begin{equation}
S = \frac{1}{\lambda}\int{ d^4\tilde{x} \left[\partial_{\mu} \tilde{\phi}^{*}   \partial^{\mu} \tilde{\phi} - \frac{1}{4} \left(\tilde{\phi}^* \tilde{\phi} - 1\right)^2\right]} 
\label{eq:action dimensionless variables}
\end{equation}
and the equations of motion are free of parameters:
\begin{equation}
\partial_{\mu} \partial^{\mu} \tilde{\phi} + \frac{1}{2} \left(\tilde{\phi}^* \tilde{\phi} - 1\right) \tilde{\phi} = 0~,
\label{eq:eoms dimensionless variables}
\end{equation}
where partial derivatives are now with respect to the dimensionless spacetime coordinates. These are the equations we solve in our simulations. In what follows, we will drop the tildes over the variables for the sake of simplicity in notation.

\subsection{Boundary conditions}
We employed periodic boundary conditions in the $z$ direction (along which the string lies) and absorbing boundary conditions in the $x$ and $y$ directions. The latter were implemented by imposing the following condition at the boundaries, for every value of the $z$ coordinate:
\begin{equation}
\frac{\partial\phi}{\partial t}+\frac{\partial\phi}{\partial x}\frac{x}{\sqrt{x^{2}+y^{2}}}+\frac{\partial\phi}{\partial y}\frac{y}{\sqrt{x^{2}+y^{2}}}\,\,\,\,\bigg\rvert_{x=\pm L_{x}/2,\,y=\pm L_{y}/2}=0\,.
\label{eq:abc}
\end{equation}
This condition is tailored to absorb radiation with radial incidence at the boundaries. Indeed, the idea behind this equation is to force the field to behave as an outgoing travelling wave at the $x,y$ boundaries. Far away from the string core, the real and imaginary parts of the field can be approximated by their vacuum expectation value plus a cylindrically symmetric travelling wave of the form
\begin{equation}
\xi\left(t,x,y\right)\propto\frac{1}{\left(x^2+y^2\right)^{1/4}}\cos\left(k\sqrt{x^{2}+y^{2}}-\beta t+\delta\right)\,,
\label{eq:travelling wave}
\end{equation}
where $k$, $\beta$ and $\delta$ are, respectively, the wave number, the angular frequency and the phase of the radiation. One can show that such a wave is a solution to Eq. (\ref{eq:abc}) if $k=\beta$, which means that the absorbing condition works better for massless radiation. However, massive radiation emitted by the string turned out to be efficiently absorbed as well.

\subsection{Details of the simulations of section IV}
For the simulation corresponding to Fig. \ref{fig:parametric-resonance-from-field-theory} in section IV, where the amplification of the zero mode occurs, we chose $L_{x}=L_{y}=80$ and $L_{z}=40$, with lattice spacing $\Delta x=0.2$ and time step $\Delta t=0.1$. The static string solution was perturbed initially with the shape mode and a small amplitude zero mode in the $x$ direction:
\begin{equation}
\phi\left(r,\theta,t=0\right)=\left[f\left(r\right)+A\left(t=0\right)\eta_{s}\left(r\right)\right]e^{i\theta}+D\left(t=0\right)\sin\left(w_{z}z\right)\eta_{x}^{0}\left(r,\theta\right)\,,
\label{eq:initial state resonance}
\end{equation}
with $A\left(t=0\right)=0.5$, $D\left(t=0\right)=0.01$ and $\omega_{z}=6\pi/L_{z}$.\\\\
For the comparison shown in Fig.~\ref{fig:comparison-theory-simulation} of the amplitudes of the modes with the solution of the coupled equations Eq. (\ref{full-analytic-eqs}), we performed a different simulation with a smaller initial amplitude of the shape mode in order to reduce the effects of radiation as much as possible. In that case we chose $A\left(t=0\right)=0.1$, $D\left(t=0\right)=0.1$ and $w_{z}=2\pi/L_{z}$, in a lattice with $L_{x}=L_{y}=80$, $L_{z}=13.8$, $\Delta x=0.1$ and $\Delta t=0.05$.\\\\
The projections needed to read the amplitude of the modes in the numerical simulations were computed as follows. Firstly, we assume that the field configuration at any time is given by 
\beq
\phi(t,{\bf x}) = f(r) e^{i \theta} + A(t) \eta_s(r)e^{i \theta} + D(t) \eta^{0}_x(r,\theta) \cos(w_z z)+R\left(t,\bf x\right)\,,
\label{eq:ansatz resonance with radiation}
\eeq
where $R(t,\bf x)$ denotes collectively the scattering states. In order to find the amplitude of the shape mode, $A(t)$, we multiply both sides of Eq. (\ref{eq:ansatz resonance with radiation}) by $\eta_{s}\left(r\right)e^{-i\theta}$ and integrate the real part over all space to get 
\begin{equation}
A\left(t\right)=\frac{1}{2\pi L_{z}}\int_{-L_{z}/2}^{L_{z}/2}dz\int_{0}^{2\pi}d\theta\int_{0}^{r_{\text{max}}}dr\,r\left[\phi_{1}\left(r,\theta,z,t\right)\cos\theta+\phi_{2}\left(r,\theta,z,t\right)\sin\theta\right]\eta_{s}\left(r\right)\,,
\label{eq:A projection}
\end{equation} 
where $\phi_{1}$ and $\phi_{2}$ are the real and imaginary parts of $\phi$, respectively. Similarly, the amplitude of the zero mode can be obtained by projecting onto $\left(\eta_{x}^{0}\right)^{*}$. The result is
\begin{equation}
D\left(t\right)=\frac{2\pi}{L_{z}}\frac{\int_{0}^{L_{z}/2}dz\int_{0}^{2\pi}d\theta\int_{0}^{r_{\text{max}}}dr\,r\left[\phi_{1}\left( f' \cos^{2}\theta+\frac{f}{r}\sin^{2}\theta\right)+\phi_{2}\left(f'-\frac{f}{r}\right)\sin\theta\cos\theta\right]}{\int_{0}^{2\pi}d\theta\int_{0}^{r_{\text{max}}}dr\,r\left[\left(	f'\right)^{2}+\left(\frac{f}{r}\right)^{2}\right]}\,.
\label{eq:D projection}
\end{equation} 
Finally, in section IV A we compute the energy radiated by the string when the parametric resonance of the zero mode takes place (see Fig. \ref{fig:power-emitted}). As mentioned in the main text, this is done by integrating the radiated power over the surface of a distant box surrounding the string. In order to account separately for the energy radiated in massive and massless modes, we follow the prescription given in \cite{Drew:2019mzc,Drew:2022iqz}. The $00$ and $0i$ components of the energy-momentum tensor can be written as
\begin{equation}
T^{00}=\Pi_{\varphi}^{2}+\left(\mathcal{D}\varphi\right)^{2}+\Pi_{\alpha}^{2}+\left(\mathcal{D}\alpha\right)^{2}+V\left(\phi_{1},\phi_{2}\right)\,,
\label{eq:Too}
\end{equation}
\begin{equation}
T^{0i}=2\left[\Pi_{\varphi}\left(\mathcal{D}\varphi\right)_{i}+\Pi_{\alpha}\left(\mathcal{D}\alpha\right)_{i}\right]\,,
\label{eq:Too}
\end{equation}
where
\begin{equation}
\Pi_{\varphi}=\frac{\phi_{1}\partial_{t}\phi_{1}+\phi_{2}\partial_{t}\phi_{2}}{\sqrt{\phi_{1}^{2}+\phi_{2}^{2}}}\,,
\label{eq:pi phi}
\end{equation}
\begin{equation}
\mathcal{D}\varphi=\left(\frac{\phi_{1}\partial_{x}\phi_{1}+\phi_{2}\partial_{x}\phi_{2}}{\sqrt{\phi_{1}^{2}+\phi_{2}^{2}}},\frac{\phi_{1}\partial_{y}\phi_{1}+\phi_{2}\partial_{y}\phi_{2}}{\sqrt{\phi_{1}^{2}+\phi_{2}^{2}}},\frac{\phi_{1}\partial_{z}\phi_{1}+\phi_{2}\partial_{z}\phi_{2}}{\sqrt{\phi_{1}^{2}+\phi_{2}^{2}}}\right)\,,
\label{eq:D phi}
\end{equation}
\begin{equation}
\Pi_{\alpha}=\frac{\phi_{1}\partial_{t}\phi_{2}-\phi_{2}\partial_{t}\phi_{1}}{\sqrt{\phi_{1}^{2}+\phi_{2}^{2}}}\,,
\label{eq:pi alpha}
\end{equation}
\begin{equation}
\mathcal{D}\alpha=\left(\frac{\phi_{1}\partial_{x}\phi_{2}-\phi_{2}\partial_{x}\phi_{1}}{\sqrt{\phi_{1}^{2}+\phi_{2}^{2}}},\frac{\phi_{1}\partial_{y}\phi_{2}-\phi_{2}\partial_{y}\phi_{1}}{\sqrt{\phi_{1}^{2}+\phi_{2}^{2}}},\frac{\phi_{1}\partial_{z}\phi_{2}-\phi_{2}\partial_{z}\phi_{1}}{\sqrt{\phi_{1}^{2}+\phi_{2}^{2}}}\right)\,.
\label{eq:D alpha}
\end{equation}
The terms with $\varphi$ correspond to the contribution of the massive modes, while those with $\alpha$ are identified as the contribution of the massless modes. Therefore, their corresponding radiated powers are given by 
\bea
P_{\text{massive}}=2 &\Big[&\int_{-L_{z}/2}^{L_{z}/2}dz\int_{-L_{y}/2}^{L_{y}/2}dy\,\left(\Pi_{\varphi}\left(\mathcal{D}\varphi\right)_{1}\big\rvert_{x=-L_{x}/2}+\Pi_{\varphi}\left(\mathcal{D}\varphi\right)_{1}\big\rvert_{x=L_{x}/2}\right)\nonumber\\
&+&\int_{-L_{z}/2}^{L_{z}/2}dz\int_{-L_{x}/2}^{L_{x}/2}dx\,\left(\Pi_{\varphi}\left(\mathcal{D}\varphi\right)_{2}\big\rvert_{y=-L_{y}/2}+\Pi_{\varphi}\left(\mathcal{D}\varphi\right)_{2}\big\rvert_{y=L_{y}/2}\right)\Big]\,,
\eea
\bea
P_{\text{massless}}=2 &\Big[&\int_{-L_{z}/2}^{L_{z}/2}dz\int_{-L_{y}/2}^{L_{y}/2}dy\,\left(\Pi_{\alpha}\left(\mathcal{D}\alpha\right)_{1}\big\rvert_{x=-L_{x}/2}+\Pi_{\alpha}\left(\mathcal{D}\alpha\right)_{1}\big\rvert_{x=L_{x}/2}\right)\nonumber\\
&+&\int_{-L_{z}/2}^{L_{z}/2}dz\int_{-L_{x}/2}^{L_{x}/2}dx\,\left(\Pi_{\alpha}\left(\mathcal{D}\alpha\right)_{2}\big\rvert_{y=-L_{y}/2}+\Pi_{\alpha}\left(\mathcal{D}\alpha\right)_{2}\big\rvert_{y=L_{y}/2}\right)\Big]\,.
\eea
Fot the simulation corresponding to Fig. \ref{fig:power-emitted} we used $L_{x}=L_{y}=80$, $L_{z}=14$, $\Delta x=0.1$ and $\Delta t=0.05$. The initial amplitude of the shape mode was $A\left(t=0\right)=0.6$ in this case.

\bibliography{parametric-instabilities-axionic-strings.bib}

\end{document}